\documentstyle[twocolumn,aps,epsfig,subeqnar,pra]{revtex}

\begin{document}
\title{Quantum phase space picture of Bose-Einstein Condensates in a double
well}
\author{Khan W. Mahmud$^{1}$\cite{khan}, Heidi Perry$^{2}$\cite{heidi}, and
William P. Reinhardt$^{1,2}$}
\address{$^{1}$Department of Physics, University of Washington, Seattle, WA
98195-1560, USA\\
}
\address{$^{2}$Department of Chemistry, University of Washington, Seattle,
WA 98195-1700, USA\\
}
\date{\today}
\maketitle

\begin{abstract}
We present a quantum phase space model of Bose-Einstein condensate (BEC) in
a double well potential. In a quantum two-mode approximation we examine the
eigenvectors and eigenvalues and find that the energy correlation diagram
indicates a transition from a delocalized to a fragmented regime. Phase
space information is extracted from the stationary quantum states using the
Husimi distribution function. We show that the mean-field phase space
characteristics of a nonrigid physical pendulum arises from the exact
quantum states, and that only 4 to 8 particles per well are needed to reach
the semiclassical limit. For a driven double well BEC, we show that the
classical chaotic dynamics is manifest in the dynamics of the quantum
states. Phase space analogy also suggests that a $\pi$ phase displaced
wavepacket put on the unstable fixed point on a separatrix bifurcates to
create a superposition of two pendulum rotor states - a macroscopic
superposition state of BEC. We show that the choice of initial barrier
height and ramping, following a $\pi$ phase imprinting on the condensate,
can be used to generate controlled entangled number states with tunable
extremity and sharpness.
\end{abstract}

\section{Introduction}

Although Bose-Einstein condensate (BEC) is well described by mean-field
theory~\cite{leggett1}, it has many aspects that can only be described in a
quantum picture containing a proper description of correlations. Examples
include number squeezing~\cite{kasevich2} and the superfluid to Mott
insulator transition~\cite{mott1} observed recently in optical lattices. The
essential underlying physics can be understood with the study of a simpler
double well BEC with a variable barrier height in the well known quantum
two-mode approximation~\cite{milburn1,spekkens1}. Quantum fluids in a double
well potential exhibit many rich phenomena related to the coherence, e.g.
the Josephson effect~\cite{josephson1} and the deBroglie wave interference
~\cite{ketterle2}. A mean-field description although appropriate in
explaining these `Josephson related effects' cannot describe the `number
squeezing effects' described earlier. In this paper we develop a quantum
phase space picture of BEC in a double well and study the connection between
the mean-field and quantum effects. As important applications of our model,
we investigate dynamics in phase space, study quantum manifestations of
classical chaos in a driven double well, and show dynamic generation of
tunable entangled number states with well defined and controlled
entanglement.

It was long ago noted by Anderson~\cite{anderson2} that the Josephson
effect, namely two quantum fluids connected by a tunnel junction~\cite
{tilley}, may be modeled as a physical pendulum. Similarly, Smerzi {\itshape
et al.} in Refs.~\cite{smerzi1} showed that the semi-classical (large N)
dynamics of two weakly linked BECs can be modeled as a classical nonrigid
physical pendulum. We begin, here, with the full quantum mechanical
description of a double well BEC in a two-mode approximation~\cite
{milburn1,spekkens1,billnatorbital}, and show that the mean-field
semiclassical limit of a nonrigid physical pendulum emerges from the exact
quantum treatment. By treating the phase and the number difference of the
condensates in two wells as conjugate variables, phase space information is
extracted from the exact (two mode) quantum wavefuntion using the Husimi
projection~\cite{husimi} of semi-classical quantum mechanics. We show that
these phase space projections of exact quantum eigenstates are localized on
the known classical energy contours of the nonrigid physical pendulum~\cite
{smerzi1}, and thus the mean-field classical phase space properties, such as
libration and $\pi $ states, are  seen to be a property of the exact quantum
eigenstates. We explore quantum classical correspondence for the stationary
states  in phase space as a function of particle number, and show that the
semiclassical limit already emerges for particle numbers as small as 4 to 8
per well. 

The quantum phase space model also reveals an underlying time dependent
semi-classical dynamics in phase space. In a study of the dynamics of a
displaced coherent state, we show a surprisingly close correspondence
between classical whorls and quantum dynamics even for N as small as 4
per well. We further illustrate that a sinusoidally driven double
well BEC (a driven physical pendulum) shows clear signatures of classical
chaos in the quantum phase space. This can be contrasted with a different
property of a chaotic system - the recently observed phenomenon of dynamical
tunneling~\cite{chaosNIST,milburnchaos}, which is a quantum motion between
two resonance zones in phase space not allowed within the classical
dynamics. We also discuss the dynamics of a coherent ground state after a
sudden change of barrier height~\cite{kasevich2,burnett2}. We show that the
oscillations between a number squeezed and a phase squeezed state is a
rotation of a pulsing ellipse in the phase space.

Due to the macroscopic nature of its wavefunction, BEC should be an ideal
system for the generation of macroscopic quantum superposition states
(Schr\"odinger cat states). The creation of macroscopic superposition states
in various condensed matter systems has received attention~\cite{friedman1}.
In the context of BEC, several authors have suggested producing such states
~\cite{cirac1,burnett1,polkovnikov1,zollercat}, although none have been
demonstrated experimentally. We show how such macroscopic quantum
superposition states are generated in phase space with a single component
BEC in a double well. Starting with a ground state centered at the origin
and displacing it through a $\pi$ phase imprinting to put it on the
hyperbolic fixed point of the classical phase space, the autonomous dynamics
splits the wavepacket along the separatrix to create entangled number states
of the form
\begin{equation}
|\Psi \rangle =\frac{1}{\sqrt{2}}\left( |n_{L},N-n_{L}\rangle
+|N-n_{L},n_{L}\rangle \right) 
\end{equation}
where $|n_{L},n_{R}\rangle $ denotes a state with $n_{L}$ particles in the
left well, $n_{R}$ in the right well, and $N=n_{L}+n_{R}$. The idea of the
exploitation of unstable fixed points to generate such entangled states with
BEC in a double well and spinor condensates in a single trap has also been
discussed in the works of Polkovnikov {\itshape et al.}~\cite{polkovnikov1}
and Micheli {\itshape et al.}~\cite{zollercat}, a discussion of which is
given in Sec.~\ref{sec:cat}. Unlike in other proposals~\cite
{cirac1,polkovnikov1,zollercat}, we use the barrier height to control the
squeezing of the initial BEC ground state, followed by a continuous change
of barrier height,  to control both the extremity (the value of $n_{L}$) and
the sharpness (the spread around $n_{L}$) of the final entangled state. A
very simple particle loss scheme~\cite{burnett1} is used here to test the
robustness of the entangled states.

The article is organized as follows. In Sec.~\ref{sec:twomode} we introduce
the model Hamiltonian, and examine its ground and excited states. In Sec.~%
\ref{sec:phasespace} we find the Husimi probablity distribution function for
the quantum states, show that the quantum states are localized on the
classical phase space orbits of a known nonrigid physical pendulum. In Sec.~%
\ref{sec:dynamics} we analyze phase space dynamics for a displaced
wavepacket, study chaotic dynamics of a driven double well and explain phase
space rotation of a ground state. In Sec.~\ref{sec:cat} we provide a phase
space analysis of the generation of tunable entangled states. Remarks and
summary in Sec.~\ref{sec:conclusion} conclude the paper.

\section{Quantum Two-State Model}

\label{sec:twomode}

\subsection{Model Hamiltonian}

The many-body Hamiltonian for a system of $N$ weakly interacting bosons in
an external potential $V({\bf r})$, in second quantization, is given by 
\begin{eqnarray}
\hat{H}=\int d{\bf r}\hat{\Psi}^{\dagger}({\bf r})\left[-\frac{\hbar^2}{2m}%
\nabla^2+V({\bf r})\right]\hat{\Psi}({\bf r})  \nonumber \\
+\frac{g}{2}\int d{\bf r}\hat{\Psi}^{\dagger}({\bf r})\hat{\Psi}^{\dagger}(%
{\bf r})\hat{\Psi}({\bf r})\hat{\Psi}({\bf r})  \label{eqn:Nhamil}
\end{eqnarray}
where $\hat{\Psi}({\bf r})$ and $\hat{\Psi}^{\dagger}({\bf r})$ are the
bosonic annihilation and creation field operators, $m$ is the particle mass,
and $g=\frac{4\pi a_s \hbar^2}{m}$ where $a_s$ is the s-wave scattering
length.

In studies of double-well BEC or two-component spinor condensates, the
low-energy many-body Hamiltonian in Eq.~(\ref{eqn:Nhamil}) can be simplified
in the well-known two-mode approximation~\cite{milburn1,spekkens1}. Many
authors have studied the double-well condensate using the two-mode
approximation. We use the model introduced by Spekkens and Sipe
~\cite{spekkens1}. The exclusion of the nonlinear tunneling terms in this model
gives rise to the Bose-Hubbard model~\cite{bosehubbard1}. The full two-mode
Hamiltonian is 
\begin{eqnarray}
&\hat{H}=\epsilon_{LL}\hat{N_{L}}+\epsilon_{RR}\hat{N_{R}}
+(\epsilon_{LR}+gT_{1}(\hat{N}-1))  \nonumber \\
&\times (a^{\dagger}_{L}a_{R}+a^{\dagger}_{R}a_{L})+\frac{gT_0}{2}(\hat{N}
^2_{L}+\hat{N}^2_{R}-\hat{N})  \nonumber \\
&+\frac{gT_2}{2}(a^{\dagger}_{L}a^{\dagger}_{L}a_{R}a_{R}+a^{\dagger}_{R}a^{
\dagger}_{R}a_{L}a_{L}+4\hat{N_{L}}\hat{N_{R}})  \label{eqn:sipeHamil}
\end{eqnarray}
where $\hat{N_{L}}=a^{\dagger}_{L}a_{L}$, $\hat{N_{R}}=a^{\dagger}_{R}a_{R}$
, $\hat{N}=\hat{N_{L}}+\hat{N_{R}}$ and 
\begin{equation}
\epsilon_{ij}=\int d{\bf r}\phi_{i}({\bf r})\left(-\frac{\hbar^2}{2m}
\nabla^2+V({\bf r})\right)\phi_{j}({\bf r})  \nonumber \\
\end{equation}
where $i,j=L,R$. 
\begin{eqnarray}
T_0=\int d{\bf r}\phi^4_{L}({\bf r}); T_1=\int d{\bf r}\phi^3_{L}({\bf r}
)\phi_{R}({\bf r});  \nonumber \\
T_2=\int d{\bf r}\phi^2_{L}({\bf r})\phi^2_{R}
\end{eqnarray}
Here $\phi_L$ and $\phi_R$ are the left and right localized single particle
Schr\"odinger wavefunctions, the $\epsilon_{LL}$ and $\epsilon_{RR}$ are the
energies of a single particle in the left and right wells, $\epsilon_{LR}$
is the single particle tunneling amplitude; $T_0$ is the mean-field energy
in each well and $T_{1,2}$ are nonlinear tunneling matrix elements.

We make a one parameter approximation~\cite{billnatorbital} of the single
particle energies and the tunneling matrix elements: 
\begin{eqnarray}
g=1;\epsilon_{LL}=\epsilon_{RR}=T_0=1;\epsilon_{LR}=T_1=-e^{-\alpha}; 
\nonumber \\
T_2=-e^{-2\alpha}.
\end{eqnarray}
This parametrization allows a simple study of continuous change in the
linear and non-linear tunneling through variation of a single parameter 
$\alpha$. In our computations with this model we ignore the $T_2$ term which
scales as $\exp{(-2\alpha)}$. The model Hamiltonian then reduces to 
\begin{eqnarray}
&\hat{H}=\epsilon_{LL}\hat{N_{L}}+\epsilon_{RR}\hat{N_{R}}
+(\epsilon_{LR}+gT_{1}(\hat{N}-1))  \nonumber \\
&\times (a^{\dagger}_{L}a_{R}+a^{\dagger}_{R}a_{L})+\frac{gT_0}{2}(\hat{N}
^2_{L}+\hat{N}^2_{R}-\hat{N})  \label{eqn:BHhamil}
\end{eqnarray}

\subsection{Fock State Analysis}

The most general state vector is a superposition of all the number states 
\begin{equation}
|\Psi\rangle=\sum_{n_L=0}^{N}c^{(i)}_{n_L}|n_L,N-n_L\rangle
\label{eqn:basis}
\end{equation}
where 
\begin{equation}
|n_L,N-n_L\rangle=\frac{(a^{\dagger}_{L})^{n_L}}{\sqrt{n_L!}}\frac{
(a^{\dagger}_{R})^{n_R}}{\sqrt{(N-n_L)!}}|vac\rangle
\end{equation}
Finding the eigenvalues and eigenvectors of the model Hamiltonian in the
Fock basis can be easily accomplished by diagonalizing a $(N+1)*(N+1)$
tridiagonal matrix.

\begin{figure}
\begin{center}
\epsfig{figure=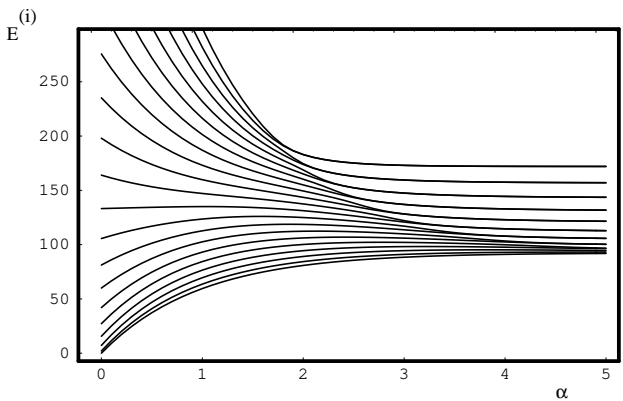,width=8.2cm}
\end{center}
\caption{Energy correlation diagram for 20 particles showing the eigenvalues as a function of barrier height $\alpha$. Note the merging of energy levels as tunneling decreases.}
\label{fig:correlation}
\end{figure}

Authors in Ref.~\cite{spekkens1} studied condensate fragmentation by looking at the ground state as the barrier is raised. We extend their analysis to look at the coefficients of the higher lying states and examine the energy correlation diagram. Fig. ~\ref{fig:correlation} shows all 21 eigenvalues for a system of 20 particles
in a double well for $\alpha$ ranging from 0 to 5. For this range of $\alpha$
, the tunneling parameters vary from 1 to 0.0067, going from a low barrier
to a high barrier leading to a fragmented condensate with fixed number of particles in each
well. The correlation diagram shows avoided crossings and energy level
merging. As $\alpha$ increases the levels start to get doubly degenerate; at
a value of about $\alpha=1.8$ the highest levels are degenerate, and all but
the ground state is degenerate for higher values of $\alpha$.

\begin{figure}
\begin{center}
\epsfig{figure=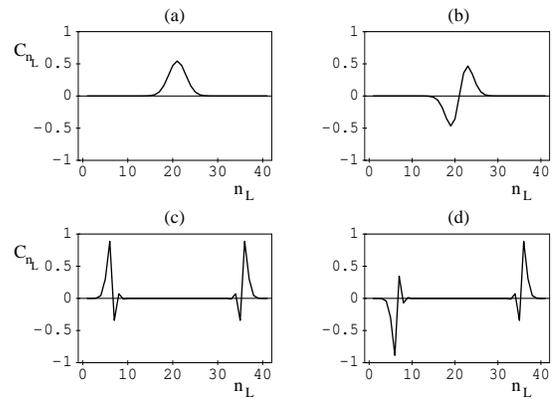,width=7.2cm}
\end{center}
\caption{Fock state coefficients for N=40 for (a) the ground state, (b) the first excited state, (c) the 30th state and (d) the 31st state. Low lying states are similar to harmonic oscillator wavefunctions, whereas the higher lying states are macroscopic quantum superpositions of particles simultaneously in both wells.}
\label{fig:coeffs}
\end{figure}

Looking at the coefficients of eigenvectors reveals interesting
characteristics of the ground and excited states. Fig.~\ref{fig:coeffs}(a)
and (b) shows the coefficients of the eigenvectors for the two lowest lying
states for 40 particles. The lowest delocalized states appear to be like the
co-ordinate space wave functions of a harmonic oscillator. These are the
states that are below the crossover ridge in a correlation diagram as in
Fig.~\ref{fig:correlation}. For states over the ridge, a similar list of
coefficients for two higher lying states are shown in Figs.~\ref{fig:coeffs}
(c) and (d). These do not look like the harmonic oscillator wave functions.
These are examples of states that are superpositions of a macroscopic number
of particles on left and right well. For these nearly degenerate
Schr\"odinger cat-like even and odd states, a very high precision arithmetic
is required to get the coefficients.

\section{Quantum Mechanical Phase Space Analysis}

\label{sec:phasespace}

\subsection{Classical Hamiltonian}

The classical Hamiltonian that describes the mean-field dynamics of BEC in a
double well has been analyzed in several papers~\cite{smerzi1,anglin1}. In a
mean-field assumption~\cite{leggett1} for the two-mode double well, and for
large enough $N$, the operators $\hat{a_j}$ can be replaced by the c-numbers 
$\sqrt{n_j}e^{i\theta_j}$ where $j=L,R$. With this assumption and defining 
$n=\frac{n_L-n_R}{2}$, $\theta=\theta_L-\theta_R$, and starting with our
model Hamiltonian Eq.~(\ref{eqn:BHhamil}) gives the classical Hamiltonian 
\begin{eqnarray}
H_{cl}=E_{c}n^2-E_{J}\sqrt{1-\left(\frac{2n}{N}\right)^2}\cos\theta+\frac{
E_{c}}{4}N^2  \nonumber \\
-\frac{E_{c}}{2}N +\epsilon_{LL}N_L+\epsilon_{RR}N_R  \label{eqn:ClassHamil}
\end{eqnarray}
where $E_{c}=gT_0$ and $E_{J}=-N\left(\epsilon_{LR}+gT_1(N-1)\right)$. Here 
$n$ and $\theta$ are conjugate variables and the equations of motion are 
\begin{subeqnarray}
\dot{n}=-E_{J}\sqrt{1-\left(\frac{2n}{N}\right)^2}\sin\theta\\
\dot{\theta}=2E_{c}n+\frac{4E_{J}n}{\sqrt{1-\left(\frac{2n}{N}\right)^2}}\cos\theta
\label{eqn:dynamics}
\end{subeqnarray}
Eq.~(\ref{eqn:ClassHamil}) is the Hamiltonian of a nonrigid physical
pendulum where $\theta$ and $n$ are the angle and angular momentum of the
pendulum. The phase space of a nonrigid physical pendulum allows novel
dynamical regimes such as the macroscopic quantum self-trapping (MQST) and 
$\pi$-motions~\cite{smerzi1}. MQST refers to the incomplete oscillations of
the populations between the two wells. $\pi$-motion refers to oscillations
such that the average relative phase remains $\pi$.

\subsection{Husimi Distribution Function}

Since the phase-space distribution function allows one to describe the
quantum aspects of a system with as much classical language allowed, it is a
popular tool to study semi-classical physics. Among the most popular
distribution functions used are the Wigner distribution, Husimi
distribution, and the Q-function~\cite{husimi,qnoise}. They are all related
- the Q-function is a special case of Husimi distribution function, and a
smoothing of the Wigner function with a squeezed Gaussian gives the Husimi
distribution~\cite{husimi}.

Husimi distribution function can be used to project, in a squeezed coherent
state representation, the classical (q,p) phase space behavior from a
stationary quantum wavefunction. Coherent state representation of the
electromagnetic field, where $n$ and $\theta$ are conjugate variables
corresponding to the number and phase of the electromagnetic fields were
introduced by Glauber~\cite{loudon}. The (q,p) coherent state
~\cite{perelomov1} is defined as 
\begin{equation}
|\beta\rangle=e^{(-|\beta|^2/2)}\sum_{n^{\prime}=0}^{\infty}\frac{
\beta^{n^{\prime}}}{\sqrt{n^{\prime}!}}|n^{\prime}\rangle
\end{equation}
which is a superposition of the harmonic oscillator eigenstates 
$|n^{\prime}\rangle$, here $\beta=q+ip$. For BEC in a double well, the phase
difference $\theta$=$\theta_L$-$\theta_R$ and the number difference $n=\frac{
n_L-n_R}{2}$ are the conjugate variable analogous to $q$ and $p$
respectively. Therefore, in ($n$,$\theta$) representations, the coordinate
and momentum representations of a squeezed coherent state is 
\begin{eqnarray}
\langle \theta^{\prime}|\theta+in\rangle=\frac{1}{(\pi\kappa)^{1/4}}
\exp\lbrack-in\theta^{\prime}-\frac{(\theta^{\prime}-\theta)^2}{2\kappa}
\rbrack \\
\langle n^{\prime}|\theta+in\rangle=\frac{1}{(\pi\kappa)^{1/4}}
\exp\lbrack-i\theta n^{\prime}-\frac{(n^{\prime}-n)^2}{2\kappa}\rbrack
\end{eqnarray}
In this representation a probability distribution function can be defined as 
\begin{equation}
P_{j}(n,\theta)=\vert\langle\theta+in|\Psi_{j}\rangle\vert^2
\end{equation}
where 
\begin{equation}
\langle\theta+in|\Psi_{j}\rangle=\frac{1}{(\pi\kappa)^{1/4}}
\sum_{n^{\prime}=-N/2}^{N/2}c^{j}_{n^{\prime}}\exp\lbrack i\theta n^{\prime}-
\frac{(n^{\prime}-n)^2}{2\kappa}\rbrack  \label{eqn:husimi}
\end{equation}
Here $n^{\prime}=\frac{n_L-n_R}{2}$, rather than being the simpler left
particle counter, and $c_{n^{\prime}}$ is the corresponding Fock-state
coefficient. Husimi function is defined for any value of the squeezing
parameter $\kappa$. The Q-function in quantum optics is a special case of
Husimi distribution function whenever $\kappa$=$\omega$, where $\omega$ is
the frequency of a coherent state Gaussian wavepacket~\cite{husimi}. The
`coarse-graining' parameter $\kappa$ determines the relative resolution in
phase space in the conjugate variables number and phase.

\subsection{Quantum Classical Connection for the Eigenstates}

\label{subsec:connection}

\begin{figure}
\begin{center}
\epsfig{figure=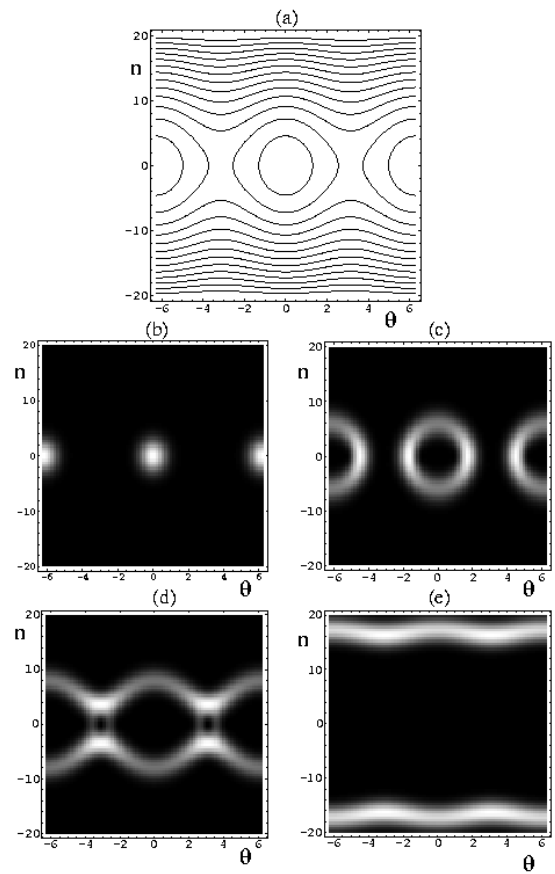,width=7.2cm}
\end{center}
\caption{Comparison of the classical nonrigid physical pendulum phase space with the Husimi distributions for different energy eigenstates for 40 particles. Shown are (a) classical energy contour. Husimi projections for (b) ground state (c) 6th (d) 12th and (e) 35th state.}
\label{fig:regularComp}
\end{figure}

It is natural to ask what aspects of the mean-field phase space properties
of a nonrigid physical pendulum~\cite{smerzi1} are contained in the exact
quantum treatment. We explore that question here by investigating the ground
and excited states of the two-mode quantum Hamiltonian, and extracting phase
space information through the use of the Husimi distribution function. Fig.
~\ref{fig:regularComp}(a) shows the classical energy contours for 40
particles for parameter values $\alpha=4$, $g=1$, $T_0=1$. For these same
parameters, Fig.~\ref{fig:regularComp} panels (b), (c), (d) and (e) show the
Husimi distributions for the ground state, 6th, 12th and 35th states
respectively. The Husimi projections confirm the physical pendulum
characteristics of the eigenstates. As is evident from the panels, the
ground state is a minimum uncertainty wave-packet in both number and phase
that is centered at the origin, the harmonic-oscillator-like low lying
excited states are the analog of pendulum librations, and the higher lying
cat-like states are the analog of pendulum rotor motions, with a clear
signature of the quantum separatrix state where the libration and rotation
states separate.

\begin{figure}
\begin{center}
\epsfig{figure=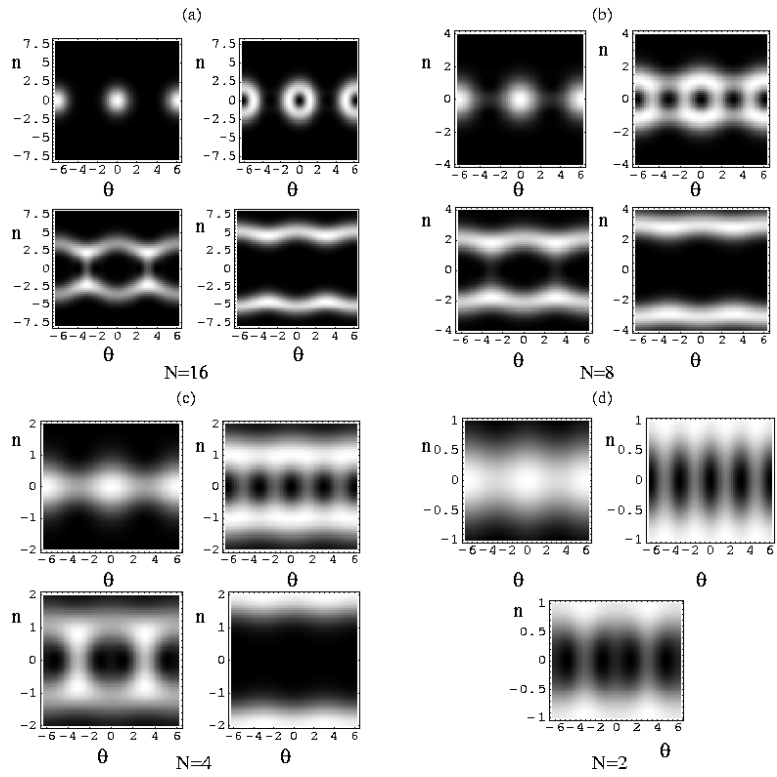,width=10.2cm}
\end{center}
\caption{Quantum-classical correspondence in phase space as functions of number of particles. Shown are the ground state, an oscillator state, a state near the separatrix and an entangled state for particle numbers (a) 16 (b) 8 (c) 4 and (d) 2. A clear signature of classical pendulum phase space is manifest for $N=8$}
\label{fig:smallN}
\end{figure}

A systematic exploration is made of the quantum classical correspondence in
phase space for different number of particles. Fig.~\ref{fig:smallN} shows
the Husimi distribution for $N=16,8,4,2$ in panels (a), (b), (c) and (d)
respectively. For each of the particle numbers, it shows the ground state, a
low lying oscillator state, a higher lying separatrix state, and a
macroscopic superposition state. Although the classical energy contours (as
shown in Fig.~\ref{fig:regularComp}(a)) are the same for all different
particle numbers, we see here that for $N=4$ and $N=2$ the minimum
uncertainty spread of the eigenstates blur the clear signature of a pendulum
phase space structure. It is interesting to note that only 4 particles per
well particles are needed to reach the semi-classical limit where the
classical phase space structure is evident. For a very large number of
particles the Husimi distributions of the eigenstates become sharper
approaching the classical limit of a line trajectory.

A fundamental difference between the classical trajectories and the quantum
states is visible in the rotor state in Fig.~\ref{fig:regularComp}(e) which
is a superposition of most particles in the left and right wells. In the
classical sense this corresponds to two different trajectories corresponding
to rotor motions of a physical pendulum in two opposite directions. The
quantum states always maintain the parity of the Hamiltonian and hence the
combinations of two such classical trajectories make up a quantum state. The
localized motion corresponding to one classical trajectory is known as
macroscopic quantum self-trapping (MQST)~\cite{smerzi1}. Such parity
violating states also appear as stationary solutions of the Gross-Pitaevskii
equation in a double well~\cite{kmahmud2}.

\begin{figure}
\begin{center}
\epsfig{figure=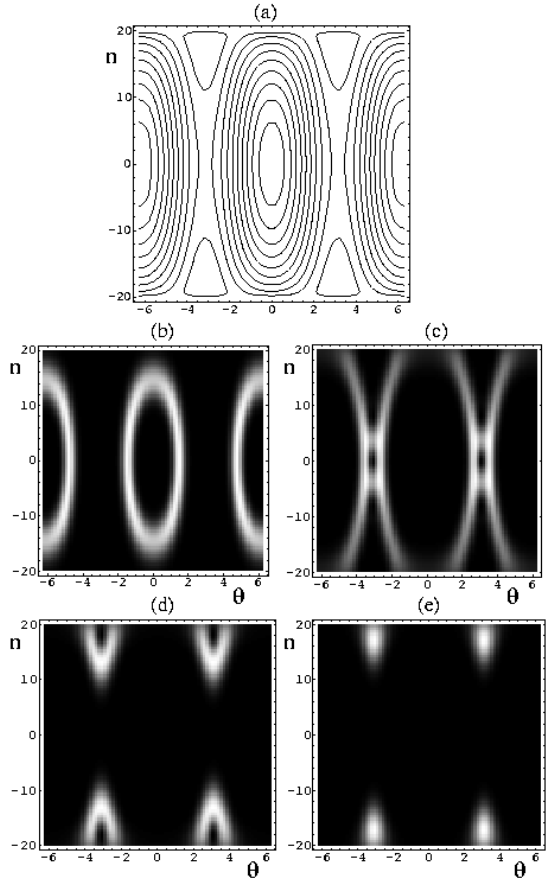,width=7.2cm}
\end{center}
\caption{Comparison of the classical and quantum phase space for N=40 showing the analog of $\pi$ states in the exact quantum treatment. Shown are (a) classical energy contour. Husimi projections are for (b) 12th (c) 30th (d) 34th and (e) 41st states. (d) and (e) are the analogs self-trapped $\pi$ states of mean-field theory, the quantum states here preserve parity.}
\label{fig:piDiamond2}
\end{figure}

In order for the quantum Hamiltonian to correspond to a momentum-shortened
physical pendulum, there should exist $\pi$ type motions~\cite{smerzi1}
among the quantum states. A change in the parameters to $\alpha=4$, $g=0.1$
and $T_0=0.1$, puts us in a slightly different regime as shown in Fig.
~\ref{fig:piDiamond2}(a) showing dynamical regimes with an average phase
difference of $\pi$. The Husimi projections in panels (b), (c), (d) and (e)
are respectively for the 12th, 30th, 34th and the 41st state. Here the
higher-lying quantum states are the analog of $\pi$-motions of the
mean-field classical Hamiltonian. Again only 4 to 8 particles per well are
needed to reach the semiclassical limit.

\section{Dynamics in phase space}

\label{sec:dynamics}

\subsection{Comparison of classical and quantum dynamics}

\label{sec:revival}

\begin{figure}
\begin{center}
\epsfig{figure=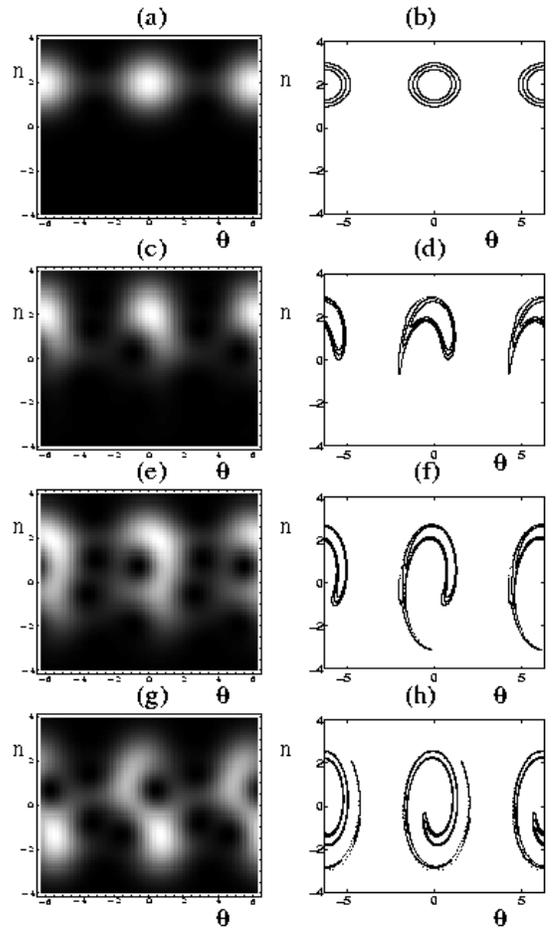,width=7.2cm}
\end{center}
\caption{A comparison of quantum and classical dynamics for N=8. We see that the classical points very closely follow the quantum phase space density. The panels are for (a) initially, and after (b) the first cycle, (c) second cycle and (d) fourth cycles. The quantum interference effects for shorter times seem to have localizing effects in the region with high density of classical whorls ((g) and (h)). For much longer times quantum dynamics shows recurrences as in the next figure.}
\label{fig:n8}
\end{figure}

To illustrate the applications of the quantum phase space picture, here we
make a comparison of the quantum and classical phase space dynamics.
Investigation of the quantum classical correspondence in phase space by
approximating a Gaussian wavepacket with a swarm of points in the classial
phase space, although widespread in quantum chaos literature, has not been
performed for BEC. This type of comparison between non-averaged quantities
contains the maximum amount of information allowed. By approximating the
quantum wavepacket with a swarm of points in the classical phase space, the
mean-field and quantum dynamics is compared for 8 particles in Fig.
~\ref{fig:n8}. The first column shows the quantum dynamics in Husimi projection
space for a $N/4$ displaced wavepacket, and the second column shows the
corresponding classical points initially, after the first, second and fourth
cycles respectively. The effects of dephasing is apparent in the quantum
phase space in panels (e) and (g). The classical trajectories develop a
narrow whorl-type structure as shown in Fig.~\ref{fig:n8}(f) and (h).
Surprisingly even for such small number of particles the classical and
quantum dynamics is comparable; the quantum states are localized in the
region of the classical points with high phase space density. For a longer
time scale the whorls become more convoluted and finer, and the quantum
dynamics shows prominent interference effects such as recurrences as
discussed next.

\begin{figure}
\begin{center}
\epsfig{figure=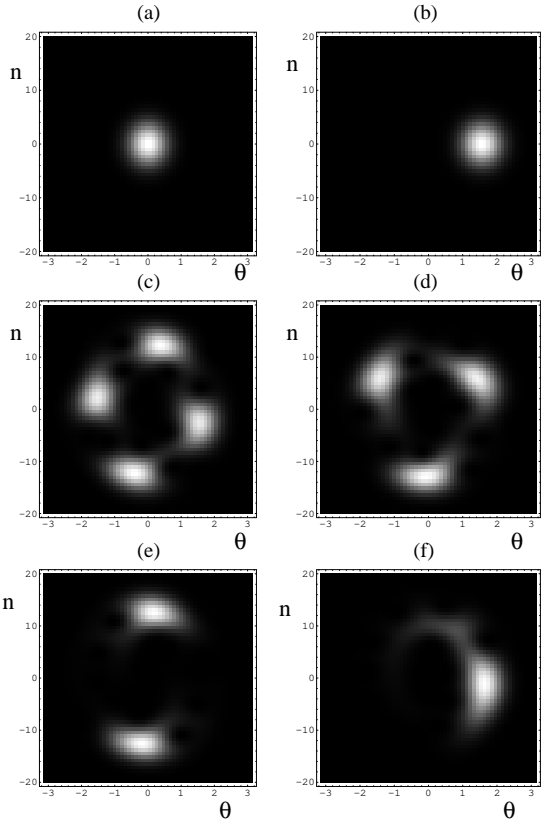,width=7.2cm}
\end{center}
\caption{Husimi projections showing fractional revivals in the dynamics of a phase displaced ground state for N=40. The recurrence time here is T=12.375. The panels show (a) ground state, (b) ground state phase displaced by $\pi/2$, and revivals approximately at (c) T/4, (d) T/3, (e) T/2 and (f) T. (e) is an example of a macroscopic superposition of two coherent states, and (f) is approximately a full revival of (b)}
\label{fig:fracRevival}
\end{figure}

Schr\"odinger~\cite{schrodinger2} first pointed out that quantum time
evolution of a displaced harmonic oscillator ground state led to a minimum
uncertainty wavepacket which evolves in time following its classical phase
space trajectory without any spreading. In the nonlinear pendulum considered
here, a ground state displaced by a small amount will evolve in phase space
without much spreading. However a state which is farther from the origin
will show the effects of nonlinearity and quantum interference a lot
quicker. After the full delocalization occurs, the interference effects
become pronounced for longer times. Localized peaks appear which again
delocalizes with the appearance of new peaks. Fig.~\ref{fig:fracRevival}
shows such fractional revivals~\cite{wright1,mott2} in the Husimi projection
space for N=40.

\subsection{Quantum-Classical correspondence for classically chaotic dynamics
}

\label{sec:chaos} 
\begin{figure}
\begin{center}
\epsfig{figure=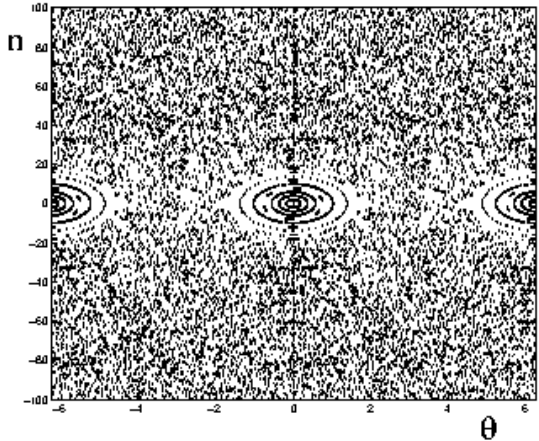,width=7.2cm}
\end{center}
\caption{A composite Poincar\'e surface of section for 100 trajectories evenly spaced on $\theta=0$. This is for N=200, and for a sinusiodal barrier $\alpha=2.5+2.5Cos(10t)$.}
\label{fig:poincare}
\end{figure}

In the context of chaotic dynamics in BEC, dynamical
tunnelling of untracold atoms from a BEC in a modulated periodic potential
has been observed~\cite{chaosNIST}, and a theoretical study of a similar
system has been done using the Floquet operator~\cite{milburnchaos}. These
authors showed that exact quantum dynamics of the
system can exhibit classically forbidden tunnelling between two regular
regions in the corresponding classical phase space, a phenomenon known as
dynamical tunneling~\cite{heller1}. Here we study instead the similarities
in the dynamics in the classical and quantum phase space. A driven pendulum
is a well known example of a one and half degree of freedom classical system
exhibiting chaos. For an analogous system of a driven double well BEC, we
make a comparison of the quantum states at different times with the
corresponding classical trajectories and illustrate signatures of quantum
chaos. Such comparison is done in phase space most usefully between the
Husimi projection of a quantum state and the corresponding classical band of
points initially in the same region of phase space~\cite{takahashi1}. For a
diagnostic to the classical phase space, Fig.~\ref{fig:poincare} shows the
Poincar\'e section for 200 particles and $\alpha=2.5+2.5cos(10 t)$. As the
amplitude of the driving force becomes larger the whole phase space becomes
chaotic.

\begin{figure}
\begin{center}
\epsfig{figure=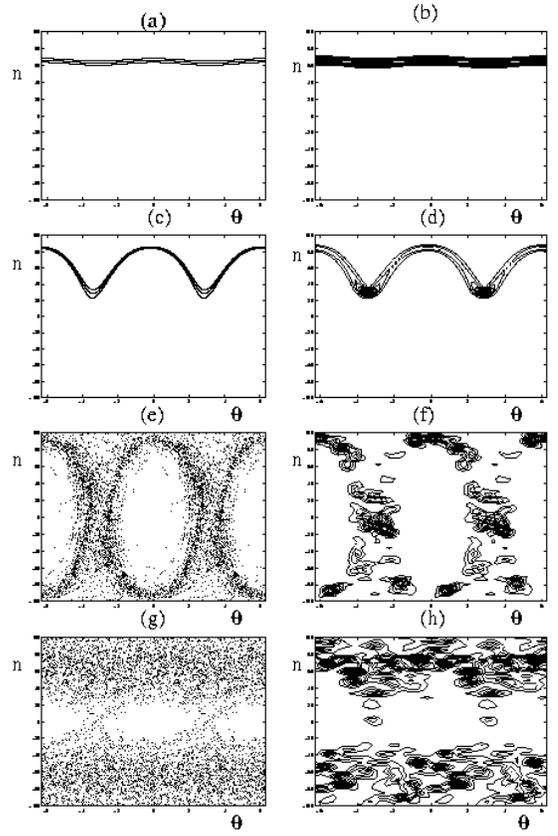,width=7.2cm}
\end{center}
\caption{Comparison of classical and quantum dynamics for points in the chaotic regions of phase space for N=200. Right panels show Husimi projections for the time evolution of the localized superposition of 128th and 129th eigenstate and the left panels show the time evolution of three bands of classical trajectories intially localized in the same region. (a),(b) at t=0 (c),(d) t=0.17 (e),(f) t=1.7 and (g),(h) t=5. Quantum states are visibly localized around the chaotic classical points.}
\label{fig:bandChaos}
\end{figure}

For comparisons in the chaotic region, the Husimi distribution of the
superpositions of 128th and 129th eigenstate at different times are shown in
Fig.~\ref{fig:bandChaos} on the right panels. The classical trajectories of
similar points are shown in the left panels. At shorter time t=0.17 as shown
in Fig.~\ref{fig:bandChaos}(c) and (d), the quantum state very nicely
follows the classical points. Panels (e) and (f) are a comparison for points
showing a visibly chaotic yet localized pattern both in the classical and
quantum phase space. The effect of chaotic dynamics fully takes effect at
t=5 when the classical phase space points are diffused throughout the whole
region as shown in (g). A comparison with the Husimi projections in (h)
makes evident the manifestations of chaos in the quantum dynamics. A state
initially localized in the regular regions of phase space does not give rise
to such chaotic structures. In the limit when $\hbar \rightarrow 0$ or
equivalently $1/N \rightarrow 0$, the discrete quantum energy spectrum
becomes continuous and the quantum mechanics will more closely follow
classical mechanics; any evidence of chaos in the quantum dynamics will be
better represented in such comparisons.

\subsection{Relative number and phase squeezing}

\label{sec:pulsing} Ground state number-squeezing with a variable barrier
height in double and multi-well systems has been discussed and observed by
many authors~\cite{kasevich2,mott1,spekkens1}. The case of a sudden change
of barrier height on a coherent ground state, which we analyze here, has been discussed on a theoretical basis~\cite{kasevich2,burnett2}. In Ref~\cite{burnett2}, the authors consider the evolution, in the space of number differences, of an intially perfect binomial number distribution state, and find that for an
optimal value of parameters in the Hamiltonian, the initial state
periodically evolves to a relatively number squeezed state.

\begin{figure}
\begin{center}
\epsfig{figure=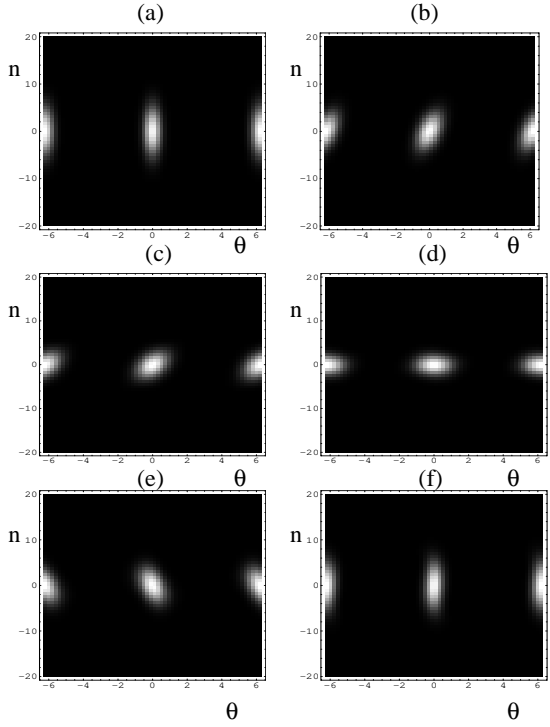,width=7.2cm}
\end{center}
\caption{Husimi projections showing rotations in phase space for N=40. (a) The initial phase squeezed or coherent state at t=0, (b) slightly rotated state at t=0.075, (c) t=0.09, (d) at t=0.125, a number squeezed state, (e) t=0.165 (f) at t=0.25 the evolution brings the state back to the initial phase squeezed state.}
\label{fig:pulsingHus}
\end{figure}

We perform here a quantum phase space analysis of this phenomenon, and find
this to be a propety of coherent ground state evolving under a Hamiltonian
for which it is not an eigenstate. We show that the initial state rotates in
 the number-angle phase space and thus becomes elongated or well defined in
number and phase periodically. We illustare this with an example: the ground
state for $\alpha =0$ very closely approximates a state with a binomial
distribution of Fock state coefficients. With a sudden raising of barrier to 
$\alpha =3$, we follow the evolution of the state in phase space. The
initial coherent ground state is not an eigenstate of the changed potential
and hence will time evolve accordingly. As shown in the quantum phase space
in Fig.~\ref{fig:pulsingHus} panels (a), the initial state is rather well
defined in phase ($\theta $) and elongated in number difference ($n$).
Further evolution in the new potential rotates the elongation in phase space
such that after a certain period it becomes well defined in $n$ (as in panel
(d)) or it is relatively number squeezed. A full cycle is shown in Fig.~\ref
{fig:pulsingHus}; in (f) the evolution brings it back to the initial
coherent state.

\section{Generating tunable entangled states using phase engineering}

\label{sec:cat}

The quantum phase space model presented here points to a simple way that an
entangled state can be generated with a single component BEC in a double
well. A wave-packet $\pi$ phase displaced to the unstable hyperbolic fixed
point of a classical phase space bifurcates along the separatrix if allowed
to time evolve. With the above motivation, here we provide a visual
explanation in phase space of the creation of controlled entangled number
states of a BEC in a double well via phase imprinting on the part of the
condensate in one of the wells followed by a continuous change of barrier
height. When properly implemented this results in a state of the form

\begin{equation}
|\Psi\rangle=\frac{1}{\sqrt{2}}\left(|n_L,N-n_L\rangle+|N-n_L,n_L\rangle
\right)
\end{equation}
where $|n_L,n_R\rangle$ denotes a state with $n_{L}$ particles in the left
well, $n_{R}$ in the right well, with total number of particles 
$N=n_{L}+n_{R}$. Unlike in other proposals~\cite
{cirac1,burnett1,polkovnikov1,zollercat}, we can use the barrier height to
control the squeezing of the initial BEC ground state followed by a
continuous change of barrier height to control both the extremity (the value
of $n_{L}$ ($n_{L}=0,1,2...N$)) and the sharpness (the spread around $n_L$)
of the entangled state. An extreme entangled state would correspond to 
$n_{L}=0$ or $N$.

Writing phases on part of a condensate is experimentally feasible via
interaction with a far off-resonance laser. This method has been used to
generate dark solitons and measure their velocities due to a phase offset
~\cite{billsoliton}. Mathematically, such a method corresponds to multiplying
the coefficient of each of the Fock states in the expansion of an eigenstate
by $e^{in_{L}\theta}$, where $|n_{L}\rangle $ is the corresponding Fock
state, and $\theta$ is the phase offset for particles in the left well. By 
$\pi$ phase imprinting the condensate in one well, the ground state centered
at the origin (0,0) in phase space is displaced to the unstable equilibrium
point (0,$\pi $) on the separatrix. Using exact quantum time evolution
within the framework of the two mode model, the resulting quantum
wave-packet bifurcates as expected. If the barrier is raised as discussed
below, the wave-packet is permanantly split, resulting in a superposition of
two classical rotor states.

\subsection{Entangled state generation without decoherence}

\begin{figure}
\begin{center}
\epsfig{figure=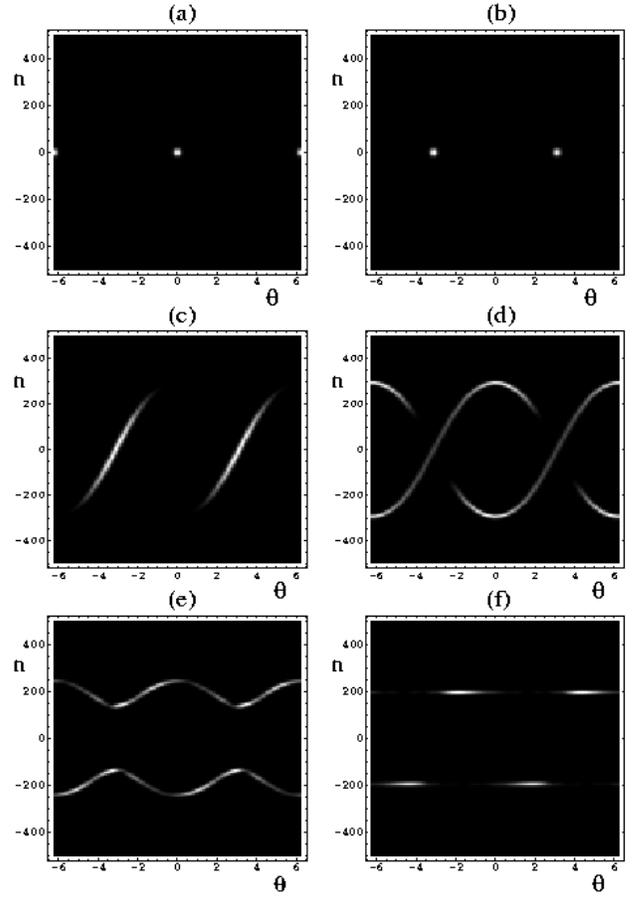,width=8.2cm}
\end{center}
\caption{Shown is the evolution to an entangled state of N=1000 in Husimi projection space. (a) The ground state at t=0, (b) the $\pi$-phase imprinted ground state at the hyperbolic fixed point, (c) at t=0.01 the wave-packet is bifurcating along the separatrix, (d) at t=0.016 it continues to move along the separatrix, (e) at t=0.4 the states become trapped as we increase the barrier, and (f) at t=2.3 a sharply peaked entangled state is obtained.}
\label{fig:cathusimiL}
\end{figure}

\begin{figure}
\begin{center}
\epsfig{figure=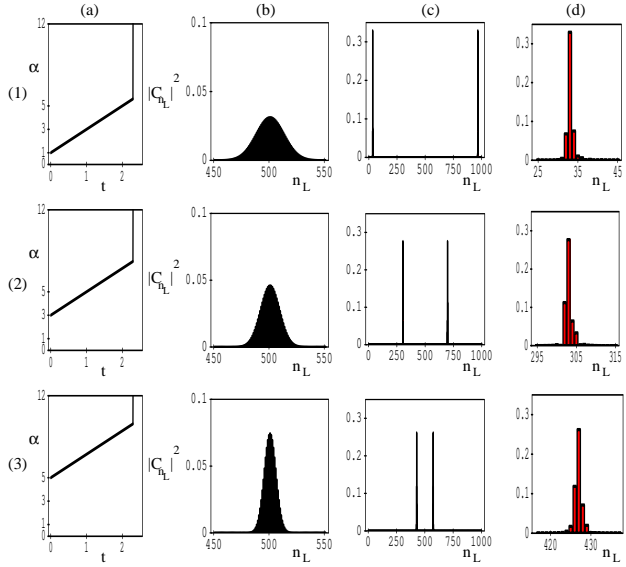,width=8.2cm}
\end{center}
\caption{Shown are the entangled states for N=1000 with different initial heights of the barrier and therefore different initial squeezings of the BEC ground state, but the same ramping of the potential. Row (1) shows the states where $\alpha=1+2t$: (a) the parameter $\alpha$ as a function of time, (b) the ground state, (c) the final entangled state, and (d) a magnified view of the Fock-state coefficients. Rows (2) and (3) show the results for $\alpha=3+2t$ and $\alpha=5+2t$ respectively. The initial barrier height controls the extremity of the entangled states. Note that for clarity the axes in the panels have different scalings.}
\label{fig:catcompareL}
\end{figure}

In the situation when there is no decoherence, well controlled entangled states
can be generated within the two-mode quantum dynamics. As an example, Figs.
~\ref{fig:cathusimiL} show how a number entangled state with 1000 particles
is generated. Fig.~\ref{fig:cathusimiL} shows the evolution in phase space
using Husimi projections - (a) the ground state, (b) a $\pi$-phase imprinted
state, (c) and (d) show subsequent evolution in the process of bifurcating
the state; further evolution along with a change of barrier totally splits
and traps the state symmetrically above the separatrix, as shown in (e),
finally giving rise to an entangled state in (f). Here the barrier height is
ramped up in time as $\alpha=3+2t$. When an entangled state is reached the
barrier is suddenly raised to essentially halt the evolution. With different
initial barrier heights and the same ramping of the potential, the extremity
of the entangled states can be tuned. Examples are shown in Fig.~\ref
{fig:catcompareL} where the different values of the barrier heights are 
$\alpha=1+2t$, $\alpha=3+2t$ and $\alpha=5+2t$ for rows (1), (2) and (3)
respectively. The columns show: (a) the barrier height and the ramping, (b)
the respective ground state, (c) the final entangled state at the end of the
ramping, and (d) a close view of the coefficients for the final state shows
that these are rather sharply peaked entangled states. As is evident from
the pictures, the initial squeezing of the ground state determines the
extremity of the final entangled state. The rate at which the barrier is
ramped determines the sharpness.

\subsection{Entangled state generation with loss}

Macroscopic superposition states are not observed mainly due to interaction
with the environment. In elastic collisions where the total number of atom
is conserved, phase damping destroys the quantum coherence~\cite{savage2}.
In the case where the number of particles are not conserved, the loss of
even a single particle destroys an extreme entangled state~\cite{burnett1},
as can be seen with the operation of a destruction operator to such a state 
\begin{equation}
\hat{a_1}(|N,0\rangle+|0,N\rangle)/\sqrt{2}=\sqrt{N/2}|N-1,0\rangle
\end{equation}

\begin{figure}
\begin{center}
\epsfig{figure=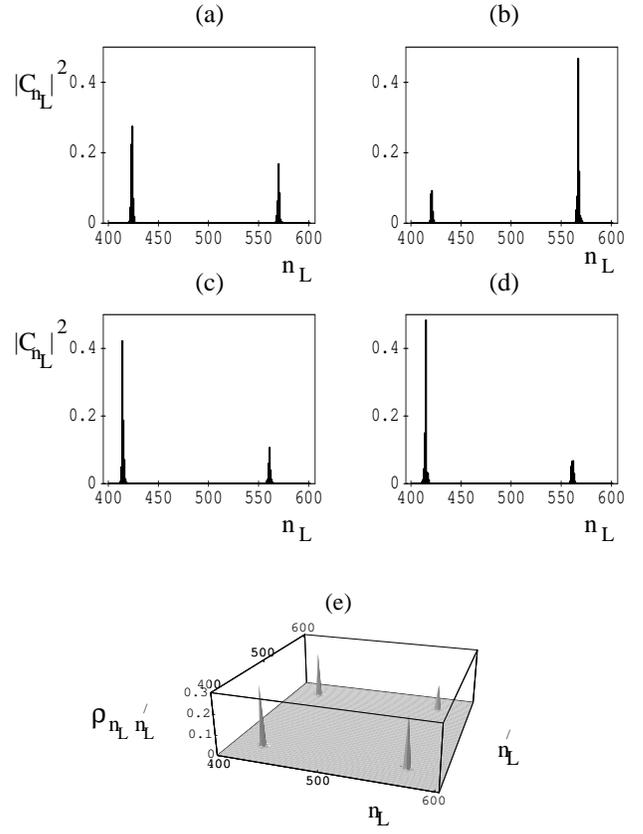,width=8.2cm}
\end{center}
\caption{Effects of loss of particles on entangled states. (a) and (b) show the effect of loss of 10 particles on the less extreme entangled state example of the third row in the previous figure. (c) and (d) show the effects of loss of 30 particles. (e) shows density matrix for panel (a) denoting that the coherence is not lost.}
\label{fig:loss}
\end{figure}

The robustness of the entangled states is tested with such a loss scheme. It
is likely that particles from the condensate will be lost during the
evolution of the state when the barrier is raised. This is simulated by the
operation of the destruction operator at different time intervals during the
evolution and taking particles out randomly from either well at each time.
Fig.~\ref{fig:loss} shows different realizations of loss of different number
of particles from the least extreme entangled state example in Fig.~\ref
{fig:catcompareL}, third row. Panels (a) and (b) are two different
simulations for a loss of 10 particles during the evolution. Panels (c) and
(d) show two different runs for a loss of 30 particles from the same
entangled state. Results for extreme entangled states are not shown here as
such states are totally destroyed, meaning all the particles are localized
in one well. The simulations suggests that a less extreme entangled state is
more robust, so it may be desirable to sacrifice the extremity of a cat
state in order for it to survive in a realistic laboratory setting. To
compare the effects of loss for sharpness, an entangled state which is not
sharp and has a Gaussian spread has a better chance of having nonvanishing
coefficients after the loss of particles. So the most robust state would be
a less extreme entangled state with a Gaussian width of coefficients around
the two peaks. The coherence is not lost in destroying particles in the
fashion done here - this is evident in the density matrix~\cite{savage2} for
panel (a) as shown in panel (e). The off-diagonal peaks in the density
matrix that quantifies the coherence remains a geometric mean of the
diagonal elements since we have not introduced phase damping; coherence
vanishes only when the final state is localized in one well.

\subsection{Discussions}

During our development of the quantum phase space picture for the double
well BEC since 2002~\cite{billnatorbital,kmahmud3}, several other authors
have also noted that metastable quantum states and dynamical instability can
be exploited to produce entangled states in a double well~\cite{polkovnikov1}
and in a spinor condensate~\cite{zollercat}. All these findings are
consistent with the phase space model introduced in this paper; our
demonstration of the tunability and sharpening of the entangled states in a
double well setting provides a useful improvement which may be important for
experimental detection and other practical purposes. The Wigner distribution
function, the Gaussian average of which is the Husimi distribution, has also
emerged as a valuable tool to the description of entangled state generation
in a spinor condensate~\cite{zollercat}.

\section{Remarks and Summary}

\label{sec:conclusion}

We have developed a quantum mechanical phase space picture of a double-well
Bose-Einstein condensate in the two-mode approximation. In a mean-field
approximation, the two-mode Hamiltonian reduces to the Hamiltonian of a
nonrigid physical pendulum. Examination of the Husimi projections of the
stationary quantum states reveals how the mean-field classical phase space
follows directly from quantum mechanics. We have found eigenstate structures
that are localized like classical oscillating states, free-rotor states and
$\pi$ states.

The Husimi probability distribution turns out to be an extremely useful tool
to study BECs in a double-well. Through its study we found unifying
connections and new insights into the double well phase space and its
dynamics. For a driven double well, quantum states are found to diffuse into
the chaotic region of phase space analogous to classical chaos. A $\pi$
phase imprinted condensate put on an unstable fixed point of the classical
phase space bifurcates along the separatrix if allowed to time evolve. The
extremity and the sharpness of the entangled states produced in this scheme
can be tuned with the initial barrier height and the appropriate ramping of
the potential. The model developed here may find applications in the studies
of other double well BEC dynamics, such as in a study of asymmetric wells,
effects of change of scattering lengths, transitions connected to avoided
crossings, topics in quantum chaos and studies of the effects of decoherence.

\acknowledgements

We would like to thank Sarah B. McKinney for discussions and computational
support and Mary Ann Leung for a critical reading of the manuscript. This
work was supported by NSF grant PHY-0140091.



\begin{references}
\bibitem[\ast]{khan} to whom correspondence should be addressed.
\end{references}

%
\onecolumn
\twocolumn
\end{document}